\documentclass[
letterpaper,prl,twocolumn,showpacs,floatfix,nofootinbib,preprintnumbers,superscriptaddress,amsmath,amssymb]{revtex4-1}
\usepackage[colorlinks=true, allcolors=blue]{hyperref}
\usepackage{graphicx}
\usepackage{dcolumn}
\usepackage{bm}
\usepackage{braket}
\usepackage{color}
\usepackage{xcolor}
\usepackage{array}
\usepackage{booktabs}
\usepackage{mathtools}

\begin{document}

\title{Angular and Kinetic Properties of Scission Neutrons within Time-dependent Density Functional Theory}

\author{Antonio Bjelčić}
\email{Corresponding author: bjelcic1@llnl.gov}
\affiliation{Lawrence Livermore National Laboratory, Livermore, California 94550, USA}

\author{Ibrahim Abdurrahman}
\affiliation{Facility for Rare Isotope Beams, Michigan State University, East Lansing, Michigan 48824, USA}

\author{Kyle Godbey}
\affiliation{Facility for Rare Isotope Beams, Michigan State University, East Lansing, Michigan 48824, USA}

\begin{abstract}
Scission-neutron emission is investigated in
$^{235}\mathrm{U}(\mathrm{n}_{\mathrm{th}},\mathrm{f})$, 
$^{239}\mathrm{Pu}(\mathrm{n}_{\mathrm{th}},\mathrm{f})$ and
$^{252}\mathrm{Cf}(\mathrm{sf})$
within time-dependent density functional theory.
Using a substantially larger simulation domain than in previous studies, the angular and energy distributions of emitted scission neutrons are extracted over a specific range of emission angles. At these angles, scission neutrons are absent below a threshold energy of roughly $1.5$--$2\,\mathrm{MeV}$, and instead contribute predominantly to the higher energy part of the prompt fission neutron spectrum. Combining the calculated scission-neutron spectrum with a Maxwellian model for the evaporated component, constrained by low-energy experimental data, reproduces the measured high-energy prompt-fission-neutron yield in both
$^{239}\mathrm{Pu}(\mathrm{n}_{\mathrm{th}},\mathrm{f})$ and
$^{252}\mathrm{Cf}(\mathrm{sf})$, whereas the evaporation-only model systematically underestimates it.
This identifies a signature of scission neutrons already present in existing high-energy prompt fission neutron spectra and constitutes direct evidence for a non-negligible scission-neutron component in prompt fission neutron emission.
\end{abstract}

\date{\today}
\maketitle


Nuclear fission was discovered by~\textcite{Hahn:1939} in 1938 and explained semi-classically by~\textcite{Meitner:1939} in 1939, yet a complete microscopic description remains lacking despite considerable recent progress~\cite{Bender:2020,Bulgac:2020,Schunck:2022}.
A predictive microscopic framework is critical for understanding stages of the process inaccessible directly by experiment, including the evolution of the compound nucleus from the outer saddle point to scission, the subsequent neck rupture, and the formation of primary fission fragments (FFs).
It is at these stages that the total kinetic energy, spins, masses, charges, and excitation energies of the primary FFs are established, which control the subsequent emission of detectable neutrons and gamma rays. 

Current statistical evaporation codes~\cite{Verbeke:2018,Talou:2021,Koning:2023} assume all prompt neutrons are emitted isotropically in the rest frames of fully accelerated FFs. In principle, other possibilities exist, such as the emission of neutrons while the FF accelerate or during the neck rupture. The latter, termed scission neutrons (SNs),  are formed in a highly non-equilibrium dynamical process and therefore lie entirely outside the reach of statistical models.
SNs were conjectured as early as 1939 by~\textcite{Anderson:1939} and~\textcite{Bohr:1939}, experimentally investigated since 1962~\cite{Bowman:1962}, and have been studied using various models for decades~\cite{Stavinsky:1959,Fuller:1962,Boneh:1974,Madler:1985,Milek:1988,Brosa:1990,Brosa:1992,Carjan:2007,Rizea:2008,Carjan:2010,Capote:2016,Carjan:2019,Randrup:2026}.
Despite this long history, SNs remain experimentally unresolved: prompt-neutron observables alone do not distinguish whether a detected neutron was emitted at scission or later evaporated from a fragment, making any extraction of a SN signal inherently model dependent.

Recently, SNs were predicted for the first time within a fully microscopic framework based on time-dependent density functional theory (TDDFT) extended to superfluid systems~\cite{Abdurrahman:2024}.
That study reported SN emission in $^{235}\mathrm{U}(\mathrm{n}_{\mathrm{th}},\mathrm{f})$, $^{239}\mathrm{Pu}(\mathrm{n}_{\mathrm{th}},\mathrm{f})$ and $^{252}\mathrm{Cf}(\mathrm{sf})$, but the simulation domain was too small to simultaneously determine the angular and kinetic properties of the SNs before they reached the boundary of the computational domain.
Here we employ a substantially larger computational domain, enabling the first extraction of correlated angular and kinetic distributions over a finite angular range. For the first time, this allows for an unambiguous and experimentally measurable signature of SNs to be extracted within a fully microscopic framework. In this study, a direct comparison is made between the density functional theory (DFT) predictions and recent existing measurements of the prompt fission neutron spectra (PFNS) for $^{239}\mathrm{Pu}(\mathrm{n}_{\mathrm{th}},\mathrm{f})$~\cite{Vorobyev2018ScissionPu239} and $^{252}\mathrm{Cf}(\mathrm{sf})$~\cite{Vorobyev2017ScissionCf252}. Importantly, unlike in the majority of other models, TDDFT does not a priori assume the existence of SNs, rather, they naturally emerge from the dynamics.   

\textit{Extraction of SN properties.} Simulations were performed within TDDFT using the axially-symmetric harmonic-oscillator solver \texttt{AxialHOHFB}~\cite{Bjelcic:2026}. We employed the SkM$^*$ energy density functional (EDF) with a mixed surface-volume pairing parameterized as in Ref.~\cite{Bjelcic:2026}. The simulation domain was taken to be an ellipsoid with semi-axes $Z_\mathrm{max}=70\,\mathrm{fm}$ and $R_\mathrm{max}=55\,\mathrm{fm}$, corresponding to a simulation volume about 3.2 times larger than that used in previous studies~\cite{Abdurrahman:2024}.

From these simulations, the SN angular and kinetic properties are extracted from the local neutron densities and currents over the time interval between scission and the moment when neutrons reach the boundary of the computational domain. Hereafter, $\rho(\mathbf r,t)$, $\tau(\mathbf r,t)$, and $\mathbf j(\mathbf r,t)$ denote the neutron number, kinetic, and current densities, respectively, and $m$ denotes the neutron mass.
The total kinetic $\mathcal{E}_{\mathrm{kin}}$, collective-flow $\mathcal{E}_{\mathrm{coll}}$, and intrinsic kinetic $\mathcal{E}_{\mathrm{int}}$ energy densities are defined as
\begin{align}
\mathcal{E}_{\mathrm{kin}}(\mathbf{r},t)
    &= \frac{\hbar^2}{2m}\,\tau(\mathbf{r},t), \qquad
\mathcal{E}_{\mathrm{coll}}(\mathbf{r},t)
    = \frac{\hbar^2}{2m}\,
       \frac{|\mathbf{j}(\mathbf{r},t)|^2}{\rho(\mathbf{r},t)},
       \notag \\
\mathcal{E}_{\mathrm{int}}(\mathbf{r},t)
    &= \mathcal{E}_{\mathrm{kin}}(\mathbf{r},t)
     - \mathcal{E}_{\mathrm{coll}}(\mathbf{r},t).
\label{eq:energy_densities}
\end{align}

An infinitesimal volume cell at position $\textbf{r}$ is identified as belonging to the SN region $\mathcal{V}_\eta^{(\mathrm{SN})}$ when the ratio of the collective-flow and intrinsic kinetic energy densities exceeds a threshold value $\eta$, i.e.
\begin{equation}
\textbf{r}\in\mathcal{V}_\eta^{(\mathrm{SN})}(t) \iff
\frac{\mathcal{E}_{\mathrm{coll}}(\mathbf r,t)}
     {\mathcal{E}_{\mathrm{int}}(\mathbf r,t)}
\ge \eta.
\label{eq:eta_criterion}
\end{equation}
This condition was selected as for free neutrons the kinetic energy is entirely collective, i.e. $\mathcal{E}_{\mathrm{kin}} = \mathcal{E}_{\mathrm{coll}}$, implying $\mathcal{E}_{\mathrm{int}} \rightarrow 0$ and $\eta \rightarrow \infty$. Due to the finite simulation domain, the SNs never fully decouple from the FFs, thus it is sufficient to select a finite $\eta$ that is reasonably large.  In this study $\eta = 3$ was selected. If $\eta$ is chosen too large there is a risk of severely underestimate the extent of emitted SNs. The robustness of the results, with respect to the threshold parameter, is discussed in further detail in the Supplemental Material~\cite{SM}.

Once the SN region has been identified, for each volume cell of $\mathcal{V}_\eta^{(\mathrm{SN})}$
the local velocity field
and
the collective-flow energy per neutron
are defined by
\begin{equation}
\mathbf v(\mathbf r,t)
=
\frac{\hbar}{m}\,
\frac{\mathbf j(\mathbf r,t)}{\rho(\mathbf r,t)},
\quad
\epsilon(\mathbf r,t)
=
\frac{\mathcal{E}_{\mathrm{coll}}(\mathbf r,t)}{\rho(\mathbf r,t)}
=
\frac{1}{2}m\,|\mathbf v(\mathbf r,t)|^2
.
\end{equation}
The corresponding SN emission angle, measured with respect to the positive $z$ axis chosen along the light FF direction, is
\begin{equation}
\theta(\mathbf r,t)
=
\arccos\!\left(
\frac{\mathbf v(\mathbf r,t)\cdot \hat{\mathbf z}}
     {|\mathbf v(\mathbf r,t)|}
\right).
\end{equation}

The total number of SNs emitted in an angular interval $[\theta_1,\theta_2]$ is given by
\begin{equation}
\label{Eq:Nsn_def}
N^{(\mathrm{SN})}_{[\theta_1,\theta_2]}(t)
=
\;\;\;\;\;\;\;\;\;\;
\int\limits_{\mathclap{\mathcal{V}_\eta^{(\mathrm{SN})}(t)\cap\{\theta(\mathbf r,t)\in[\theta_1,\theta_2]\}}}
\;\;\;\;\;\;\;\;\;\;
d^3\mathbf r\,
\rho(\mathbf r,t),
\end{equation}
and the SN kinetic energy distribution by
\begin{equation}
\label{Eq:Psi_sn_definition}
\Phi^{(\mathrm{SN})}_{[\theta_1,\theta_2]}(E;t)
=
\;\;\;\;\;\;\;\;\;\;
\int\limits_{\mathclap{\mathcal{V}_\eta^{(\mathrm{SN})}(t)\cap\{\theta(\mathbf r,t)\in[\theta_1,\theta_2]\}}}
\;\;\;\;\;\;\;\;\;\;
d^3\mathbf r\,
\rho(\mathbf r,t)\,
\delta\!\left(E-\epsilon(\mathbf r,t)\right).
\end{equation}

Both the number and kinetic distribution of the SNs are time-dependent. However, at sufficiently late times, they are expected to approach asymptotic time-independent limits. This is precisely why such a large simulation volume is crucial: these time-dependent properties must reach their asymptotic limits before the SNs reach the boundary of the computational domain and introduce numerical artifacts.

\textit{Angular range.}
With the present simulation domain, SN properties can be extracted reliably only for emission angles in the range $110^\circ \leq \theta \leq 150^\circ$. For other angles, the computed SN yield does not saturate before reaching the boundary of the simulation domain, see Supplemental Material~\cite{SM}. Figure~\ref{Fig:snapshots_reduced} shows the post-scission evolution of the neutron density for the $^{239}\mathrm{Pu}(\mathrm{n}_{\mathrm{th}},\mathrm{f})$ case and indicates the angular window over which reliable SN properties are extracted.

\begin{figure}
  \centering
  \includegraphics[width=0.68\columnwidth]{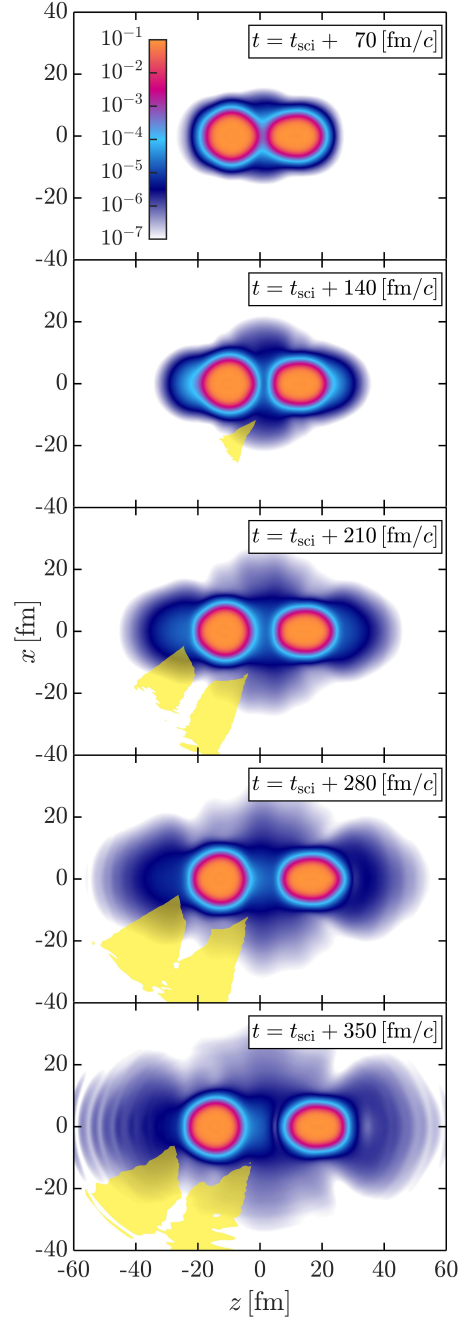}
  \caption{Snapshots of the post-scission neutron density evolution of the compound system $^{240}\mathrm{Pu}^*$.
The colormap used is the same as in Ref.~\cite{Abdurrahman:2024}.
Yellow shaded region marks the SN-occupied volume $\mathcal{V}_{\eta=3}^{(\mathrm{SN})}(t)$ corresponding to emission angles $110^\circ \leq \theta(\textbf{r},t)\leq 150^\circ$, where SN properties can be reliably extracted.
Boundary-reflection effect is visible in the final snapshot.}
  \label{Fig:snapshots_reduced}
\end{figure}



\begin{figure}
  \centering
\includegraphics[width=0.88\columnwidth]{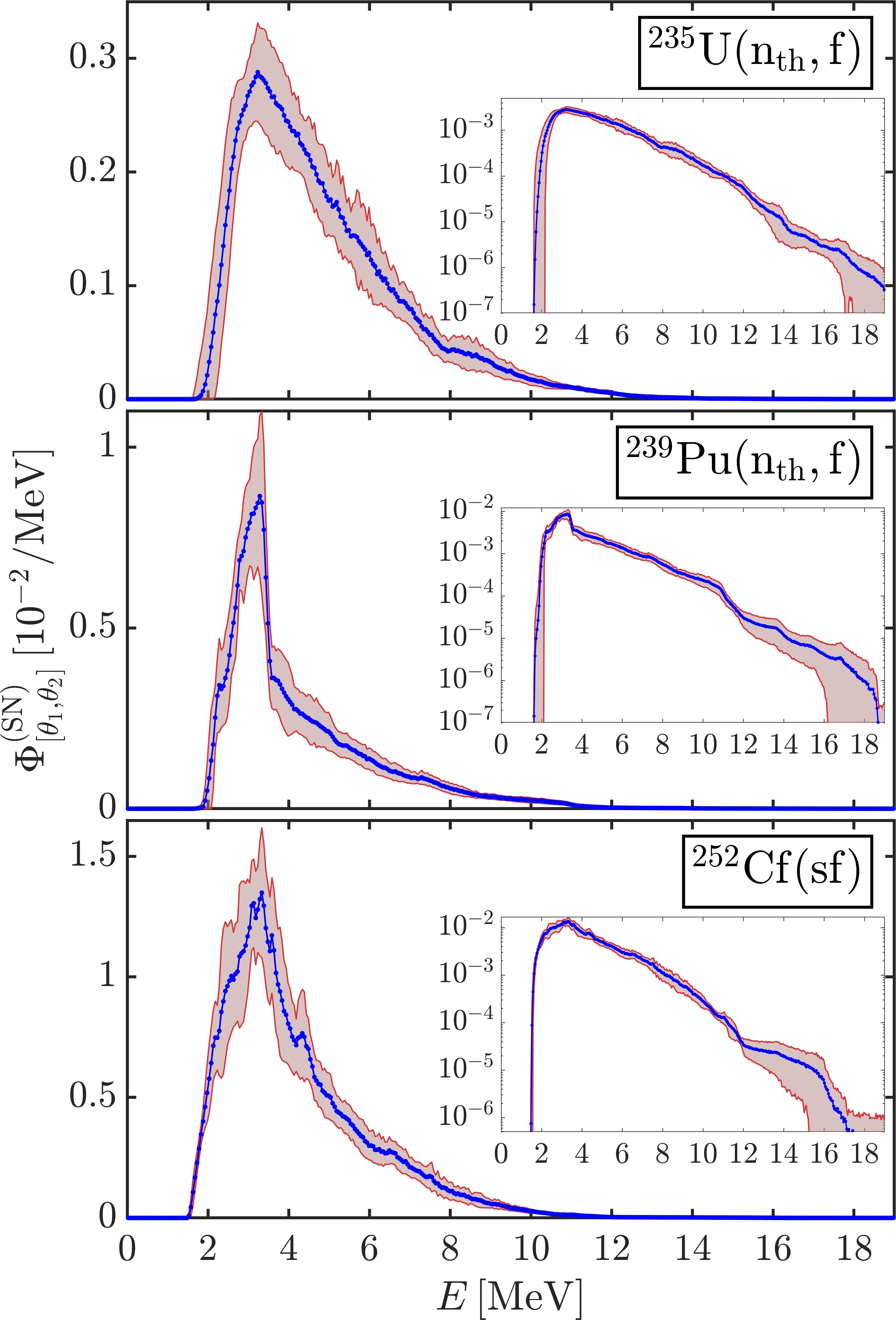}
  \caption{
SN kinetic energy distributions for three fissioning systems, computed with parameter $\eta=3$ over the angular interval
$[\theta_1,\theta_2]=[107.8^\circ , 143.2^\circ]$.
Shaded bands show one standard deviation from the time average.
Insets show the same spectra on a logarithmic scale.
For all systems, the SN spectrum
peaks near
$3$--$3.5\,\mathrm{MeV}$
and
vanishes below a threshold energy $E_\mathrm{thr}$ lying in the $1.5$--$2\,\mathrm{MeV}$ range.
    }
  \label{Fig:Phisn_combined}
\end{figure}

\textit{SN kinetic energy spectra.}
Figure~\ref{Fig:Phisn_combined} shows the SN kinetic energy distributions for $^{235}\mathrm{U}(\mathrm{n}_{\mathrm{th}},\mathrm{f})$, $^{239}\mathrm{Pu}(\mathrm{n}_{\mathrm{th}},\mathrm{f})$ and $^{252}\mathrm{Cf}(\mathrm{sf})$, all restricted to the angular interval $[\theta_1, \theta_2] = [107.8^\circ, 143.2^\circ]$.
The bands denote one standard deviation obtained from the
time-averaged distributions.

The spectra share several qualitative features across all three systems: an absence of SNs below a finite energy threshold $E_\mathrm{thr}$, a peak near $3$--$3.5\,\mathrm{MeV}$, and an approximately exponential high-energy tail extending to roughly $18\,\mathrm{MeV}$.
The cutoff occurs at $E_{\mathrm{thr}}\simeq 2\,\mathrm{MeV}$ for induced fission cases
$^{235}\mathrm{U}(\mathrm{n}_{\mathrm{th}},\mathrm{f})$ and
$^{239}\mathrm{Pu}(\mathrm{n}_{\mathrm{th}},\mathrm{f})$, compared with
$E_{\mathrm{thr}}\simeq 1.5\,\mathrm{MeV}$ for
$^{252}\mathrm{Cf}(\mathrm{sf})$, consistent with the smaller available energy in the case of spontaneous fission.

For $^{239}\mathrm{Pu}(\mathrm{n}_{\mathrm{th}},\mathrm{f})$, we obtain
a total SN yield of
$N^{(\mathrm{SN})}_{\mathrm{tot}} = 0.19 \pm 0.01$, corresponding to
approximately $6.6\%$ of the total prompt-neutron multiplicity
$\bar{\nu}=2.878\pm0.013$~\cite{Carlson2018NeutronStandards}.
For $^{252}\mathrm{Cf}(\mathrm{sf})$, we find
$N^{(\mathrm{SN})}_{\mathrm{tot}} = 0.39 \pm 0.02$, or approximately
$10.4\%$ of $\bar{\nu}=3.7610\pm0.0051$~\cite{Vorobyev2017ScissionCf252}.
Both yields are slightly smaller than those reported in
Ref.~\cite{Abdurrahman:2024}, possibly reflecting the use of a different EDF.

\textit{Evaporation model.}
The evaporated-neutron (EN) contribution to the PFNS was modeled by a Maxwellian evaporation spectrum in the rest frame of each FF, boosted to the laboratory frame using nonrelativistic kinematics.
The model contains two parameters: a common FF temperature, $T=T_\mathrm{L}=T_\mathrm{H}$, and the light-to-heavy FF neutron-multiplicity ratio, $\bar{\nu}_\mathrm{L}/\bar{\nu}_\mathrm{H}$.
Although the FFs generally emerge at different temperatures, separate values of $T_\mathrm{L}$ and $T_\mathrm{H}$ in the model do not significantly alter the EN energy spectrum, which, in the restricted angular range considered here, is dominated by neutrons evaporated from the heavy FF.

The EN kinetic energy distribution, $\Phi^{(\mathrm{EN})}_{[\theta_1,\theta_2]}(E)$, is fit to experimental data at energies below the threshold energy, $E \le E_\mathrm{thr}$: the lowest energy, for a given range of angles, at which a SN can be emitted. The total PFNS is then
\begin{equation}
    \Phi_{[\theta_1,\theta_2]}(E)
    =
    \Phi^{(\mathrm{EN})}_{[\theta_1,\theta_2]}(E)
    +
    \Phi^{(\mathrm{SN})}_{[\theta_1,\theta_2]}(E).
\end{equation} 
Because the SN contribution $\Phi^{(\mathrm{SN})}_{[\theta_1,\theta_2]}(E)$, shown in Fig.~\ref{Fig:Phisn_combined}, vanishes below $E_\mathrm{thr}$, the low-energy fit is insensitive to the SN component, and any improvement at high energies when the SN term is added is a genuine prediction of the theory. Further technical details are provided in the Supplemental Material~\cite{SM}.

\begin{figure}[t]
  \centering
  \includegraphics[width=1.00\columnwidth]{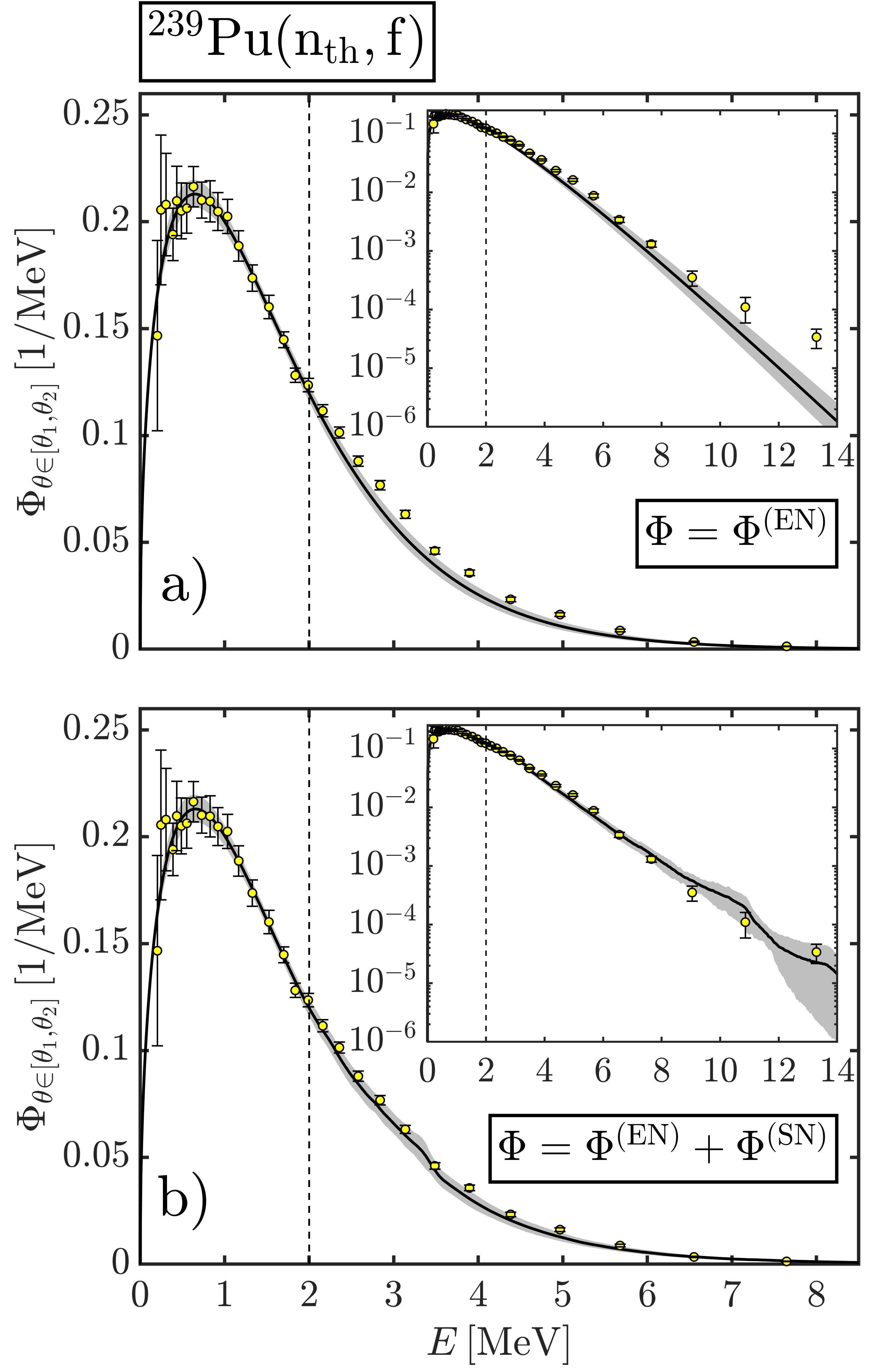}
\caption{
The PFNS for $^{239}\mathrm{Pu}(\mathrm{n}_{\mathrm{th}},\mathrm{f})$ in the angular interval $[\theta_1,\theta_2]=[107.8^\circ,143.2^\circ]$.
Experimental data are taken from Ref.~\cite{Vorobyev2018ScissionPu239}.
The upper panel, a), shows the evaporation-only model, $\Phi=\Phi^{(\mathrm{EN})}$, while the lower panel, b), shows the model including SNs, $\Phi=\Phi^{(\mathrm{EN})}+\Phi^{(\mathrm{SN})}$.
In both cases, the evaporation component is fitted only to data below $E_\mathrm{thr}=2\,\mathrm{MeV}$, indicated by the vertical dashed line.
Shaded bands denote 95\% confidence intervals.
Insets show the same spectra on a logarithmic scale.
}
  \label{Fig:Phi_Pu239}
\end{figure}

\begin{figure}[t]
  \centering
  \includegraphics[width=1.00\columnwidth]{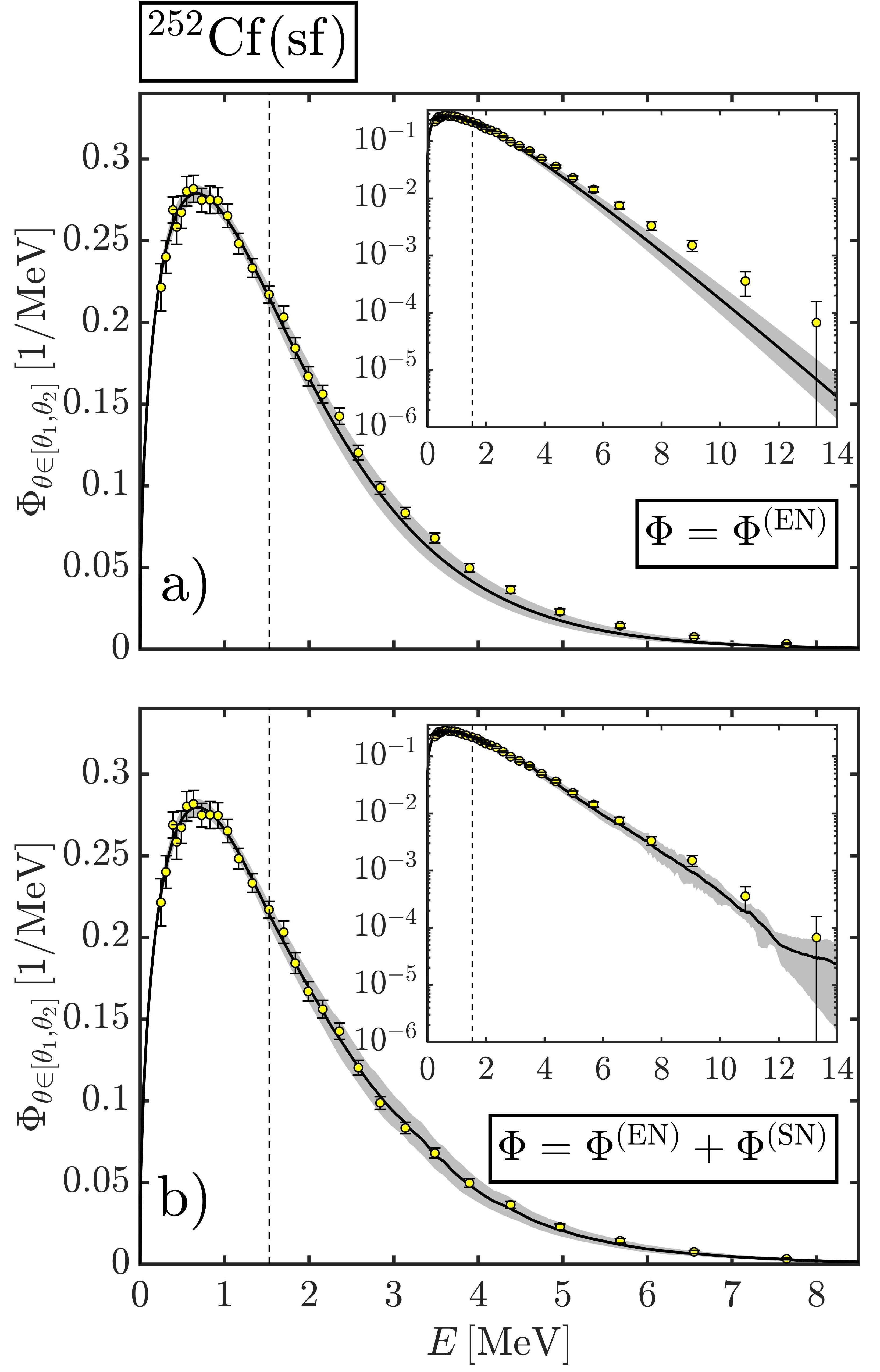}
  \caption{
  Same as Fig.~\ref{Fig:Phi_Pu239}, but for $^{252}\mathrm{Cf}(\mathrm{sf})$. Experimental data are taken from Ref.~\cite{Vorobyev2017ScissionCf252}, and the threshold energy is $E_\mathrm{thr}=1.53\,\mathrm{MeV}$.
   }
  \label{Fig:Phi_Cf252}
\end{figure}

\textit{Comparison with experiment.} 
Figures~\ref{Fig:Phi_Pu239} and~\ref{Fig:Phi_Cf252} show the total PFNS for $^{239}\mathrm{Pu}(\mathrm{n}_{\mathrm{th}},\mathrm{f})$ and $^{252}\mathrm{Cf}(\mathrm{sf})$, with and without the SN contribution. For each system, two curves are shown: one obtained using the evaporation model alone (upper panel) and one including the SN component (lower panel).
When the SN component is included, the fitting constraint is modified from $\bar{\nu}_\mathrm{L}+\bar{\nu}_\mathrm{H}=\bar{\nu}$ to $\bar{\nu}_\mathrm{L}+\bar{\nu}_\mathrm{H}=\bar{\nu}-N_\mathrm{tot}^{(\mathrm{SN})}$. In both cases, the evaporation component is fitted exclusively to low-energy experimental data, $E \le E_\mathrm{thr}$, from Refs.~\cite{Vorobyev2018ScissionPu239,Vorobyev2017ScissionCf252}. The experimental PFNS are reported at 11 discrete angles; for comparison with the microscopic prediction, we select only the data measured at $\theta=107.8^\circ$, $125.5^\circ$, and $143.2^\circ$. The best-fit evaporation-model parameters are listed in Table~\ref{tab:evap}.

\begin{table}[t]
\caption{ The best-fit parameters of the evaporation model fitted on the low-energy PFNS data from~\cite{Vorobyev2018ScissionPu239,Vorobyev2017ScissionCf252}. The evaporation model was assumed to be Maxwellian in the rest frame of the FFs. The table lists the fissioning system, if SNs were included, the FF temperature ($T$), and light-to-heavy fragment neutron multiplicity ratio ($\bar{\nu}_\mathrm{L}/\bar{\nu}_\mathrm{H}$). }
\label{tab:evap}
\begin{ruledtabular}
\begin{tabular}{l|c|c|c}
System
& SNs & $T$ [MeV] 
& $\bar{\nu}_\mathrm{L}/\bar{\nu}_\mathrm{H}$  \\ \hline
$^{239}\mathrm{Pu}(\mathrm{n}_{\mathrm{th}},\mathrm{f})$ &
No & $0.80 \pm 0.03$
& $1.56 \pm 0.03$ \\
&
Yes & $0.78 \pm 0.03$
& $ 1.39 \pm 0.03$ \\
$^{252}\mathrm{Cf}(\mathrm{sf})$
& No
& $0.84 \pm 0.03$ 
& $1.46 \pm 0.06$ \\
& Yes
& $ 0.80 \pm 0.03$
& $ 1.21 \pm 0.05$ \\
\end{tabular}
\end{ruledtabular}
\end{table}

In both systems, the model that neglects SNs systematically underestimates the measured high-energy yield, even within the 95\% confidence interval, whereas inclusion of the DFT SN contribution visibly improves the description across the full spectrum above $E_\mathrm{thr}$.


The main result of the present study is that for angular interval where reliable asymptotic properties can be extracted, in this case between the region in front of the heavy FF and the region perpendicular to the fission axis, the microscopic SN spectrum vanishes below a finite energy threshold: $2\,\mathrm{MeV}$ for both $^{235}\mathrm{U}(\mathrm{n}_{\mathrm{th}},\mathrm{f})$ and $^{239}\mathrm{Pu}(\mathrm{n}_{\mathrm{th}},\mathrm{f})$, and $1.5\,\mathrm{MeV}$ for $^{252}\mathrm{Cf}(\mathrm{sf})$. Therefore, TDDFT predicts that all prompt fission neutrons in this angular range, and below these threshold energies, are purely evaporated neutrons, that are completely decoupled from the SN contribution. 

Counterintuitively, this separation implies experimental measurements of prompt fission neutron angle-energy distributions at low neutron kinetic energies, below the threshold energies listed above, provide the best manner in which to constrain the SN contribution. Measurements for a broader set of fissioning systems are also needed: beyond the relatively recent data for $^{239}\mathrm{Pu}(\mathrm{n}_{\mathrm{th}},\mathrm{f})$~\cite{Vorobyev2018ScissionPu239} and $^{252}\mathrm{Cf}(\mathrm{sf})$~\cite{Vorobyev2017ScissionCf252}, the only other systems with published angle-energy neutron distributions are the much older $^{235}\mathrm{U}(\mathrm{n}_{\mathrm{th}},\mathrm{f})$ studies~\cite{SKARSVAG196372,OldMeasurements1,OldMeasurements2,OldMeasurements3}, which are characterized by substantial statistical noise and very limited low-energy resolution. New experimental campaigns pushing towards high precision measurements with a large angular coverage would help further validate these predictions.

On the theory side, larger simulation domains or absorbing boundary conditions~\cite{Higdon:1986,Nakatsukasa:2021,Ueda:2003,Nakatsukasa:2005} are needed to extend the accessible post-scission evolution time and to extract SN properties for all, or at least a larger range of, emission angles.
Additionally, more initial compound configurations should be considered in order to improve the robustness of the results, although the present study already represents a substantial computational effort.

In summary, this study reports the first identification of a SN signature within the existing measured PFNS, on the basis of a fully microscopic framework. For both $^{239}\mathrm{Pu}(\mathrm{n}_{\mathrm{th}},\mathrm{f})$ and $^{252}\mathrm{Cf}(\mathrm{sf})$, the discrepancy between the evaporation-only model and the experimental data, in the high-energy tail, is resolved once the microscopic SN contribution is included, even though the evaporation component is constrained solely by low-energy data that is insensitive to SNs within the specified angular range. This constitutes direct evidence for a non-negligible scission-neutron component in prompt fission neutron emission, and establishes a microscopic foundation for interpreting fission neutron spectra beyond the standard statistical-evaporation picture for the first time since SN were conjectured to exist nearly a century ago.

\begin{acknowledgments}
A.~B. acknowledges helpful discussions with N.~Schunck and M.~Verriere.
A.~B. also thanks A.~P.~Tonchev for drawing attention to relevant experimental references.
This work was partly performed under the auspices of the US Department of Energy by the Lawrence Livermore National Laboratory under Contract DE-AC52-07NA27344.
This material is partially based upon work supported by the US Department of Energy, Office of Science, Office of Advanced Scientific Computing Research and Office of Nuclear Physics, Scientific Discovery through Advanced Computing (SciDAC) program.
This work was supported in part by the U.S. Department of Energy under Award No. DOE-DE-NA0004074 (NNSA, the Stewardship Science Academic Alliances program).
Computing support came from the Lawrence Livermore National Laboratory Institutional Computing Grand Challenge program.
\end{acknowledgments}

\bibliography{zotero_output,books,others}

\end{document}


\title{Supplemental Material: Angular and kinetic properties of scission neutrons within time-dependent density functional theory}

\author{Antonio Bjelčić}
\email{Corresponding author: bjelcic1@llnl.gov}
\affiliation{Lawrence Livermore National Laboratory, Livermore, California 94550, USA}

\author{Ibrahim Abdurrahman}
\affiliation{Facility for Rare Isotope Beams, Michigan State University, East Lansing, Michigan 48824, USA}

\author{Kyle Godbey}
\affiliation{Facility for Rare Isotope Beams, Michigan State University, East Lansing, Michigan 48824, USA}

\date{\today}
\maketitle

This document provides details and supplementary figures supporting the results presented in the main manuscript.
It is organized as follows.
Section~\ref{Sec:SM_basis} describes the harmonic-oscillator basis and its spatial support properties.
Section~\ref{Sec:SM_integration} describes the two-step time integration procedure and documents the accuracy of the basis transformation.
Section~\ref{Sec:SM_extraction} describes the procedure for extracting the SN angular and energy distributions.
Section~\ref{Sec:SM_evaporation} summarizes the evaporated-neutron spectrum model and its transformation to the laboratory frame.
Section~\ref{Sec:SM_angular_stability} presents the angular stability analysis for $^{239}\mathrm{Pu}(\mathrm{n}_{\mathrm{th}},\mathrm{f})$ and $^{252}\mathrm{Cf}(\mathrm{sf})$.
Section~\ref{Sec:SM_Pu240} presents supplementary results for $^{239}\mathrm{Pu}(\mathrm{n}_{\mathrm{th}},\mathrm{f})$.
Section~\ref{Sec:SM_Cf252} presents supplementary results for $^{252}\mathrm{Cf}(\mathrm{sf})$.
Section~\ref{Sec:SM_expdata} describes the experimental data processing.

\section{Harmonic-oscillator basis\label{Sec:SM_basis}}

The Bogoliubov spinors $U(\mathbf{r},\sigma)$ and $V(\mathbf{r},\sigma)$ are expanded in an axially-symmetric harmonic-oscillator (HO) basis. The spatial HO basis functions
\begin{equation}
\Phi_{n_z,n_r,\Lambda}^{(b_z,b_r)}(\textbf{r}) = 
\phi_{n_z}^{(b_z)}(z)\,
\phi_{n_r,\Lambda}^{(b_r)}(r_\perp)\,
\frac{1}{\sqrt{2\pi}} e^{i\Lambda\varphi}
\end{equation}
are characterized by quantum numbers $(n_z, n_r, \Lambda)$ with $n_z\geq0$, $n_r\geq 0$, and oscillator lengths $b_z$, $b_r$.
Explicit definitions of $\phi_{n_z}^{(b_z)}(z)$ and $\phi_{n_r,\Lambda}^{(b_r)}(r_\perp)$ are given in Ref.~\cite{Bjelcic:2026}.

The truncated basis $\textbf{B}(N_{\mathrm{sh}}, Z_{\mathrm{max}}, R_{\mathrm{max}})$ is defined by
\begin{equation}
    \textbf{B}(N_{\mathrm{sh}}, Z_{\mathrm{max}}, R_{\mathrm{max}})
    =
    \left\{
    \Phi_{n_z,n_r,\Lambda}^{(b_z,b_r)}
    \,\middle|\,
    n_z + 2n_r + |\Lambda| \leq N_{\mathrm{sh}}
    \right\},
\end{equation}
with oscillator lengths
\begin{equation}\label{Eq:SM_bzbr}
    b_z = \frac{Z_{\mathrm{max}}}{\sqrt{2N_{\mathrm{sh}}}},
    \qquad
    b_r = \frac{R_{\mathrm{max}}}{\sqrt{2N_{\mathrm{sh}}}}.
\end{equation}
Spin \( \sigma \) and isospin \( q \) quantum numbers are implicitly included.
From the properties of Talmi--Moshinsky brackets~\cite{bjelcic2020implementation}, it can be shown~\cite{Bjelcic:2026} that the one-body local density $\rho^{(q)}(z,r_\perp)$ has an expansion in terms of HO functions whose supports satisfy
\begin{equation}
    \frac{z^2}{Z_{\mathrm{max}}^2} 
    +
     \frac{r_\perp^2}{R_{\mathrm{max}}^2}
    \leq 1.
\end{equation}
Analogously, it can be shown that all local densities and currents entering the Skyrme energy density functional are effectively contained within the ellipsoid
\begin{equation}
    \Omega = \left\{
    \mathbf{r}\in\mathbb{R}^3
    \;\middle|\;
    \frac{z^2}{Z_{\mathrm{max}}^2}
    +
    \frac{r_\perp^2}{R_{\mathrm{max}}^2}
    \leq 1
    \right\}.
\end{equation}
Increasing $N_{\mathrm{sh}}$ at fixed $Z_{\mathrm{max}}$ and $R_{\mathrm{max}}$ reduces the oscillator lengths~\eqref{Eq:SM_bzbr} and thereby improves the spatial resolution of the expansion.

\section{Two-step time integration\label{Sec:SM_integration}}

In the TDDFT study of Ref.~\cite{Abdurrahman:2024}, the Bogoliubov spinors were represented on a three-dimensional grid in a box
$2X_\mathrm{max} \times 2Y_\mathrm{max} \times 2Z_\mathrm{max} = 48 \times 48 \times 120\,\mathrm{fm}^3$,
which permitted reliable simulation of SN emission for less than $100\,\mathrm{fm}/c$ after scission.
To overcome this limitation we employ a two-step procedure.

\textit{Step 1.}
The static constrained HFB equation is solved and used as an initial condition for the TDHFB equation
\begin{equation}
    i\hbar\,\partial_t
    \left[\begin{matrix}
        U(t) \\ V(t)
    \end{matrix}\right]
    =
    \mathcal{H}\bigl(U(t),V(t)\bigr)
    \left[\begin{matrix}
        U(t) \\ V(t)
    \end{matrix}\right],
\end{equation}
which is evolved from $t=0$ to the scission moment $t=t_\mathrm{sci}$ in the basis $\textbf{B}_1$ ($N_\mathrm{sh}=70$, semi-axes $Z_\mathrm{max}=35\,\mathrm{fm}$, $R_\mathrm{max}=30\,\mathrm{fm}$, comprising 62,196 distinct $(n_z,n_r,\Lambda)$ combinations).

\textit{Step 2.}
The Bogoliubov matrices at $t=t_\mathrm{sci}$ are transformed from basis $\textbf{B}_1$ to the larger basis $\textbf{B}_2$ ($N_\mathrm{sh}=100$, semi-axes $Z_\mathrm{max}=70\,\mathrm{fm}$, $R_\mathrm{max}=55\,\mathrm{fm}$, comprising 176,851 combinations) as:
\begin{equation}
    U_{k_2,\mu}
    =
    \sum_{k_1 \in \textbf{B}_1}
    T_{k_2,k_1}\, U_{k_1,\mu},
    \qquad
    V_{k_2,\mu}
    =
    \sum_{k_1 \in \textbf{B}_1}
    T_{k_2,k_1}\, V_{k_1,\mu},
\end{equation}
via the transformation matrix:
\begin{equation}
    T_{k_2,k_1}
    =
    \delta_{\sigma_1,\sigma_2}
    \int d\mathbf{r}\,
    \left[
        \Phi_{n_{z_2}, n_{r_2}, \Lambda_2}^{(b_z^2, b_r^2)}(\mathbf{r})
    \right]^*
    \Phi_{n_{z_1}, n_{r_1}, \Lambda_1}^{(b_z^1, b_r^1)}(\mathbf{r}),
\end{equation}
for all $k_1\in\mathbf{B}_1$ and $k_2\in\mathbf{B}_2$, where $k=(n_z,n_r,\Lambda,\sigma)$.
The evolution is then continued in basis $\textbf{B}_2$ until $t=t_\mathrm{ref}$, when SN boundary reflections begin.

Time integration uses the fourth-order Runge--Kutta method with $\Delta t = 0.4\,\mathrm{fm}/c$ in Step 1 and $\Delta t = 0.2\,\mathrm{fm}/c$ in Step 2. Convergence with respect to time-step size is documented in Ref.~\cite{Bjelcic:2026}.

\textit{Basis transformation accuracy.}
Table~\ref{tab:SM_basis_conversion} summarizes the accuracy of the basis transformation for $^{240}\mathrm{Pu}^*$ at $t=t_\mathrm{sci}$.
The relative differences of order $10^{-5}$ in total energy and $10^{-6}$ in particle numbers are comparable to the numerical errors accumulated during time integration.
The quality of the transformation for other fissioning systems considered is essentially identical.

\begin{table}[t]
\caption{Comparison of the $^{240}\mathrm{Pu}^*$ total HFB energy $E_\mathrm{tot}$ and particle numbers $N$ and $Z$, evaluated using Bogoliubov spinors in bases $\textbf{B}_1$ and $\textbf{B}_2$ at $t=t_\mathrm{sci}$.}
\label{tab:SM_basis_conversion}
\begin{ruledtabular}
\begin{tabular}{l|c|c|c|c}
Quantity
& $\textbf{B}_1$
& $\textbf{B}_2$
& Abs. diff.
& Rel. diff. \\ \hline
$E_\mathrm{tot}\,[\mathrm{MeV}]$
& $-1800.3204$
& $-1800.1875$
& $0.1329$
& $7.4\times 10^{-5}$ \\
$N$
& $145.99816$
& $145.99789$
& $0.00027$
& $1.8\times 10^{-6}$ \\
$Z$
& $93.99960$
& $93.99941$
& $0.00019$
& $2.1\times 10^{-6}$ \\
\end{tabular}
\end{ruledtabular}
\end{table}

\section{Extraction of SN angular and kinetic properties\label{Sec:SM_extraction}}

The key local quantities are the neutron particle density $\rho(\mathbf{r},t)$, kinetic density $\tau(\mathbf{r},t)$, and current density $\mathbf{j}(\mathbf{r},t)$, from which we define
\begin{align}
    \mathcal{E}_{\mathrm{kin}}(\mathbf r,t)
    &=
    \frac{\hbar^2}{2m}\,\tau(\mathbf r,t),
    \\
    \mathcal{E}_{\mathrm{coll}}(\mathbf r,t)
    &=
    \frac{\hbar^2}{2m}\,
    \frac{|\mathbf j(\mathbf r,t)|^2}{\rho(\mathbf r,t)},
    \\
    \mathcal{E}_{\mathrm{int}}(\mathbf r,t)
    &=
    \mathcal{E}_\mathrm{kin}(\textbf{r},t)-\mathcal{E}_\mathrm{coll}(\textbf{r},t),
\end{align}
and the local velocity field $\mathbf{v}(\mathbf{r},t) = (\hbar/m)\,\mathbf{j}(\textbf{r},t)/\rho(\textbf{r},t)$.
The emission angle is
\begin{equation}
    \theta(\mathbf r,t)
    =
    \arccos\!\left(
    \frac{\mathbf v(\mathbf r,t)\cdot \hat{\mathbf z}}
         {|\mathbf v(\mathbf r,t)|}
    \right).
\end{equation}
A volume cell at position $\mathbf{r}$ is assigned to the scission-neutron region $\mathcal{V}_\eta^{(\mathrm{SN})}(t)\subset \Omega$ when
\begin{equation}
    \frac{\mathcal{E}_\mathrm{coll}(\mathbf r,t)}
         {\mathcal{E}_\mathrm{int}(\mathbf r,t)}
    \ge \eta,
\end{equation}
where $\eta > 0$ is a threshold parameter.
The choice $\eta \gg 1$ ensures that identified cells are dominated by outward collective flow rather than intrinsic nuclear motion; choosing $\eta$ too large underestimates the extent of $\mathcal{V}_\eta^{(\mathrm{SN})}(t)$ because the finite simulation domain prevents full decoupling.

For cells in $\mathcal{V}_\eta^{(\mathrm{SN})}(t)$, the collective-flow kinetic energy per neutron is $\epsilon(\mathbf{r},t) = \frac{1}{2}m|\mathbf{v}(\mathbf{r},t)|^2$.
The energy distribution of SN emitted in the angular interval $[\theta_1, \theta_2]$ is
\begin{equation}
\Phi^{(\mathrm{SN})}_{[\theta_1,\theta_2]}(E;t)
=
\;\;\;\;\;\;\;\;\;\;
\int\limits_{\mathclap{\mathcal{V}_\eta^{(\mathrm{SN})}(t)\cap\{\theta(\mathbf r,t)\in[\theta_1,\theta_2]\}}}
\;\;\;\;\;\;\;\;\;\;
d^3\mathbf r\,
\rho(\mathbf r,t)\,
\delta\!\left(E-\epsilon(\mathbf r,t)\right).
\end{equation}
and the total emitted number is
\begin{equation}
N^{(\mathrm{SN})}_{[\theta_1,\theta_2]}(t)
=
\int_{\mathcal{V}^{(\mathrm{SN})}_\eta(t)\cap\{\theta(\textbf{r},t)\in[\theta_1,\theta_2]\}}
d^3\mathbf{r}\,\rho(\mathbf{r},t).
\end{equation}
In practical calculations, 
the probability distribution $P^{(\mathrm{SN})}_{[\theta_1,\theta_2]} = \Phi^{(\mathrm{SN})}_{[\theta_1,\theta_2]} / N^{(\mathrm{SN})}_{[\theta_1,\theta_2]}$ is calculated using the histogram method by counting how many SNs flowing at angles $\theta(\textbf{r},t)\in[\theta_1,\theta_2]$ inside the volume $\mathcal{V}_\eta^{(\mathrm{SN})}(t)$  will have the energy $\epsilon(\textbf{r},t)$ within a given energy bin.

The obtained probability distribution $P^{(\mathrm{SN})}_{[\theta_1,\theta_2]}(E;t)$ is averaged over the finite time interval $[t_1,t_2]$ to reduce fluctuations, producing the final distribution
\begin{equation}
    \Phi^{(\mathrm{SN})}_{[\theta_1,\theta_2]}(E)
    =
    N^{(\mathrm{SN})}_{[\theta_1,\theta_2]}\,
    P^{(\mathrm{SN})}_{[\theta_1,\theta_2]}(E).
\end{equation}

\section{Evaporated-neutron spectrum model\label{Sec:SM_evaporation}}

Evaporated neutrons (ENs) are modeled by Maxwellian spectrum with neutrons emitted isotropically in the rest frame of each fragment $\mathrm{f} \in \{\mathrm{L},\mathrm{H}\}$:
\begin{equation}\label{Eq:SM_Maxwellian}
    \psi_\mathrm{f}(\varepsilon_\mathrm{cm},\theta_\mathrm{cm}) =
    \frac{\bar{\nu}_\mathrm{f}}{4\pi}
    \,
    \frac{2}{\sqrt{\pi T_\mathrm{f}^{3}}}
    \,
    \sqrt{\varepsilon_\mathrm{cm}}
    \exp\!\left(-\frac{\varepsilon_\mathrm{cm}}{T_\mathrm{f}}\right),
\end{equation}
and is normalized to $\bar{\nu}_\mathrm{f}$ neutrons per fragment
\begin{equation}
\bar{\nu}_\mathrm{f}=
\int_0^\infty d\varepsilon_\mathrm{cm} \int_0^\pi 2\pi \sin\theta_\mathrm{cm} d\theta_\mathrm{cm}\;
     \psi_\mathrm{f}(\varepsilon_\mathrm{cm},\theta_\mathrm{cm}).
\end{equation}
The Weisskopf form, $\psi_\mathrm{f} \propto \varepsilon_\mathrm{cm} e^{-\varepsilon_\mathrm{cm}/T_\mathrm{f}}$, was tested but it was found to produce a substantially poorer description of the data in both considered systems, hence all results in the main manuscript use the Maxwellian form~\eqref{Eq:SM_Maxwellian}.
This is consistent with Ref.~\cite{Vorobyev2018ScissionPu239},
where in their two-component fit,
approximately 86\% of the yield is attributed to the
Maxwellian component, and the remaining 14\% to the
Weisskopf component.

Using nonrelativistic kinematics, the rest-frame energy $\varepsilon_\mathrm{cm}$ and angle $\theta_\mathrm{cm}$ of an EN are related to its laboratory-frame energy $E$ and angle $\theta$ by
\begin{equation}
\varepsilon_\mathrm{cm} =
E + \overline{E}_\mathrm{f} - 2\sqrt{E \overline{E}_\mathrm{f}} \cos\theta,
\end{equation}
\begin{equation}
\cos\theta_\mathrm{cm} =
\frac{\sqrt{E}\cos\theta - \sqrt{\overline{E}_\mathrm{f}}}
{\sqrt{E + \overline{E}_\mathrm{f} - 2\sqrt{E \overline{E}_\mathrm{f}}\cos\theta}},
\end{equation}
where $\overline{E}_\mathrm{f} = \frac{1}{2}m\overline{V}_\mathrm{f}^2$ is the average kinetic energy per nucleon of the fully accelerated fragment moving with velocity $\overline{V}_\mathrm{f}$.
The corresponding laboratory-frame distribution for fragment $\mathrm{f}$ is
\begin{equation}
    n_\mathrm{f}(E,\theta)
    = \sqrt{\frac{E}{\varepsilon_\mathrm{cm}}}\,
    \psi_\mathrm{f}(\varepsilon_\mathrm{cm},\theta_\mathrm{cm}),
\end{equation}
and the angle-integrated evaporated spectrum in $[\theta_1,\theta_2]$ is
\begin{equation}
    \Phi_{[\theta_1,\theta_2]}^{(\mathrm{EN})}(E)
    =
    2\pi \int_{\theta_1}^{\theta_2} \sin\theta\, d\theta
    \left[
        n_\mathrm{L}(E,\theta)
        +
        n_\mathrm{H}(E,\pi-\theta)
    \right],
\end{equation}
where we adopt the convention that the light fragment moves along $+z$.
The values of $\overline{E}_L$ and $\overline{E}_H$ used in each case are taken directly from the corresponding experimental references~\cite{Vorobyev2018ScissionPu239,Vorobyev2017ScissionCf252}.

The fit parameters $T=T_\mathrm{L}=T_\mathrm{H}$ and $\bar{\nu}_\mathrm{L}/\bar{\nu}_\mathrm{H}$ are determined by minimizing
\begin{equation}\label{Eq:chi2}
    \chi^2 =
    \sum_{E_i\leq E_\mathrm{thr}}
    \left(
    \frac{\Phi^{(\mathrm{EN})}_{[\theta_1,\theta_2]}(E_i)-\Phi^{(i)}_{[\theta_1,\theta_2]}}
         {\sigma^{(i)}_{[\theta_1,\theta_2]}}
    \right)^2
\end{equation}
using experimental PFNS data $(E_i,\Phi^{(i)}_{[\theta_1,\theta_2]}\pm \sigma^{(i)}_{[\theta_1,\theta_2]})$ only below the SN threshold $E_\mathrm{thr}$.

Parameter uncertainties and 95\% confidence bands are estimated by parametric bootstrap with $10^5$ synthetic datasets, sampling all quantities with finite uncertainties ($\Phi^{(i)}_{[\theta_1,\theta_2]}$, $\bar{\nu}$, $N^{(\mathrm{SN})}_\mathrm{tot}$, and $\Phi^{(\mathrm{SN})}_{[\theta_1,\theta_2]}(E)$) from normal distributions.

An evaporation model with three fitting parameters ($\bar{\nu}_\mathrm{L}/\bar{\nu}_\mathrm{H}$, $T_\mathrm{L}$ and $T_\mathrm{H}$ separately) was also tested. The resulting fits are virtually identical to those with a single temperature $T$, because the considered angular interval $107.8^\circ \le \theta \le 143.2^\circ$ causes most light-fragment neutrons to be boosted toward $0^\circ$ in the laboratory frame, leaving $T_\mathrm{L}$ largely unconstrained and with high uncertainty. Therefore, for simplicity, $T = T_\mathrm{L} = T_\mathrm{H}$ is taken throughout the study.

\section{Angular stability analysis\label{Sec:SM_angular_stability}}

\begin{figure}
    \centering
    \includegraphics[width=1.00\columnwidth]{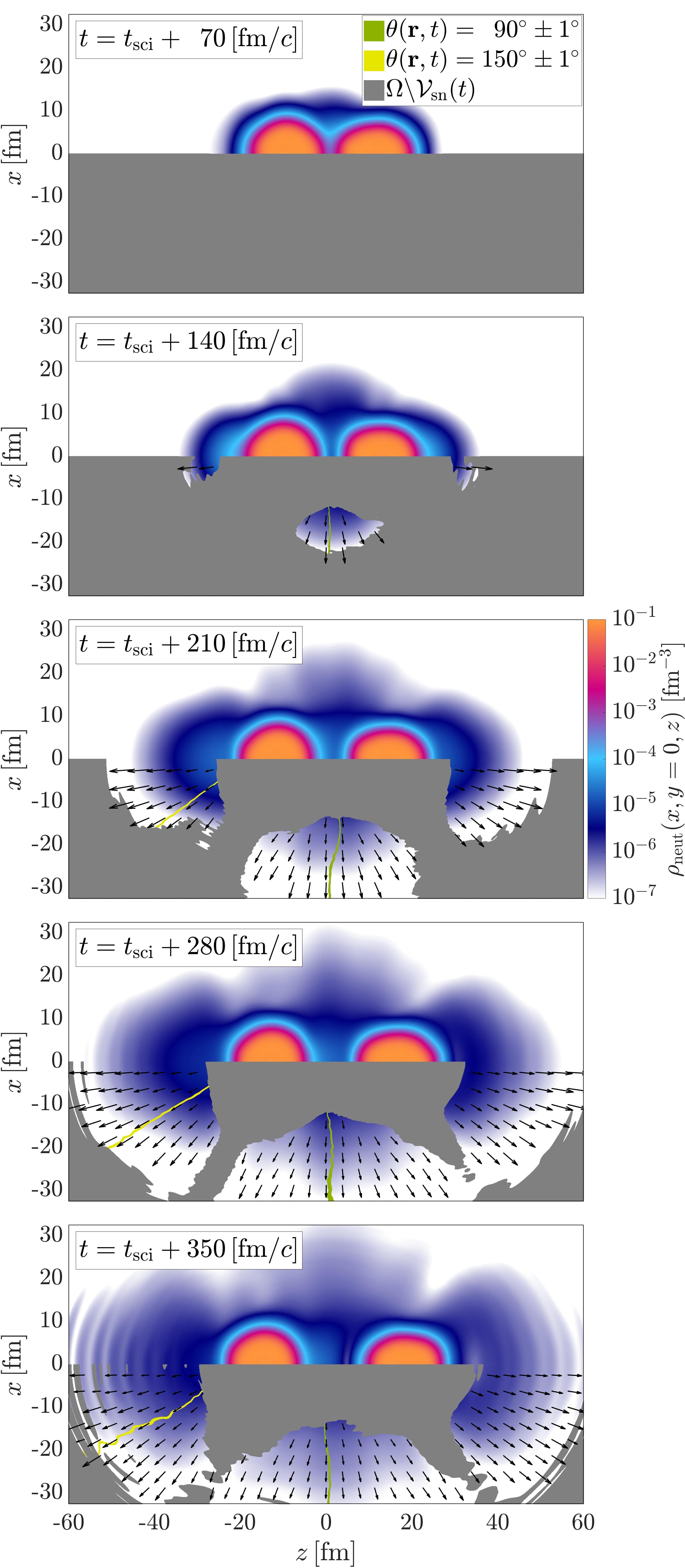}
    \caption{Snapshots of the $^{240}\mathrm{Pu}$ fission trajectory at $70\,\mathrm{fm}/c$ intervals, from $t=t_\mathrm{sci}+70\,\mathrm{fm}/c$ to $t=t_\mathrm{ref}=2350\,\mathrm{fm}/c$ are shown.
    The upper half-space contains the neutron density, with the same color map as in Ref.~\cite{Abdurrahman:2024}. The lower half-space contains the scission-neutron volume $\mathcal{V}_\eta^{(\mathrm{SN})}(t)$ for $\eta=3$, with the shaded gray space indicating the excluded region. The two colored rays indicate SNs flowing at angles $\theta(\textbf{r},t)\in[89^\circ,91^\circ]$ and $\theta(\textbf{r},t)\in[149^\circ,151^\circ]$.
    The arrows show the velocity field $\mathbf{v}(\mathbf{r},t)$.}
    \label{Fig:SM_dens_snapshots}
\end{figure}

Fig.~\ref{Fig:SM_dens_snapshots} shows the full sequence of post-scission density snapshots for $^{239}\mathrm{Pu}(\mathrm{n}_{\mathrm{th}},\mathrm{f})$.
Boundary reflections are visible in the final panel at $t=t_\mathrm{ref}$, where reflected and outgoing SNs interfere and artificially distort the extracted energies and yields.
SNs propagating at angles of $90^\circ\pm 1^\circ$ and $150^\circ\pm 1^\circ$ are also shown with green and yellow lines respectively. There is a pronounced qualitative difference between the two cases: the $150^\circ$ trajectories are approximately ray-like, whereas the $90^\circ$ trajectories exhibit significantly more curvature.


\begin{figure}
    \centering
    \includegraphics[width=1.00\columnwidth]{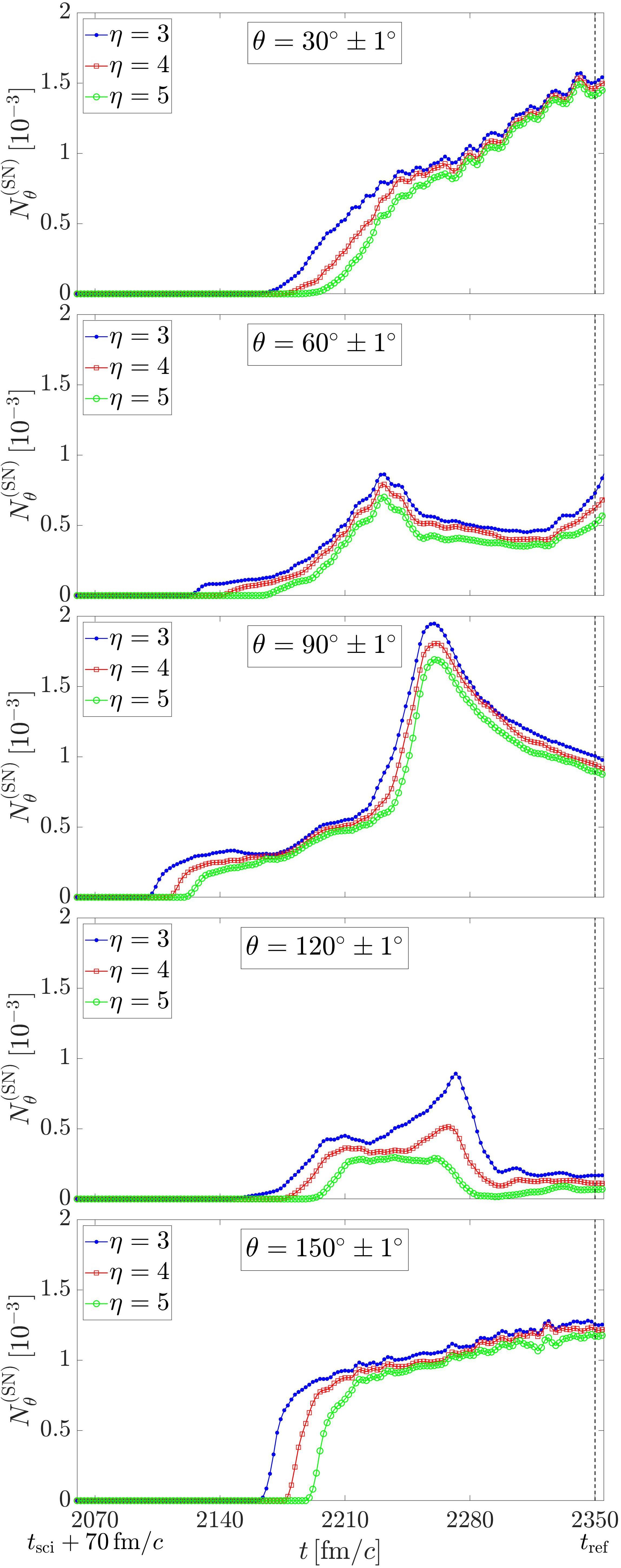}
    \caption{The time dependence of $N_\theta^{(\mathrm{SN})}(t)$ for angular windows centered at $\theta = 30^\circ$, $60^\circ$, $90^\circ$, $120^\circ$, $150^\circ$ (width $\pm1^\circ$), is shown for $\eta = 3$, 4, 5.
    Clear saturation before $t_\mathrm{ref}$ is observed only for $\theta = 120^\circ$ and $150^\circ$.}
    \label{Fig:SM_neut_time}
\end{figure}

Fig.~\ref{Fig:SM_neut_time} shows the time evolution of the SN yield at five discrete angles.
Only for $\theta = 120^\circ \pm 1^\circ$ and $\theta = 150^\circ \pm 1^\circ$ does $N_\theta^{(\mathrm{SN})}(t)$ saturate before $t_\mathrm{ref}$, confirming that reliable asymptotic yields can be extracted only in the range $110^\circ \lesssim \theta \lesssim 150^\circ$. For other emission angles, an even larger simulation box is needed.
The variation of the parameter $\eta$ mainly changes the overall magnitude of $N_\theta^{(\mathrm{SN})}(t)$, while leaving its shape in time essentially unchanged,
indicating the robustness with respect to $\eta$.

\begin{figure}
    \centering
    \includegraphics[width=0.95\columnwidth]{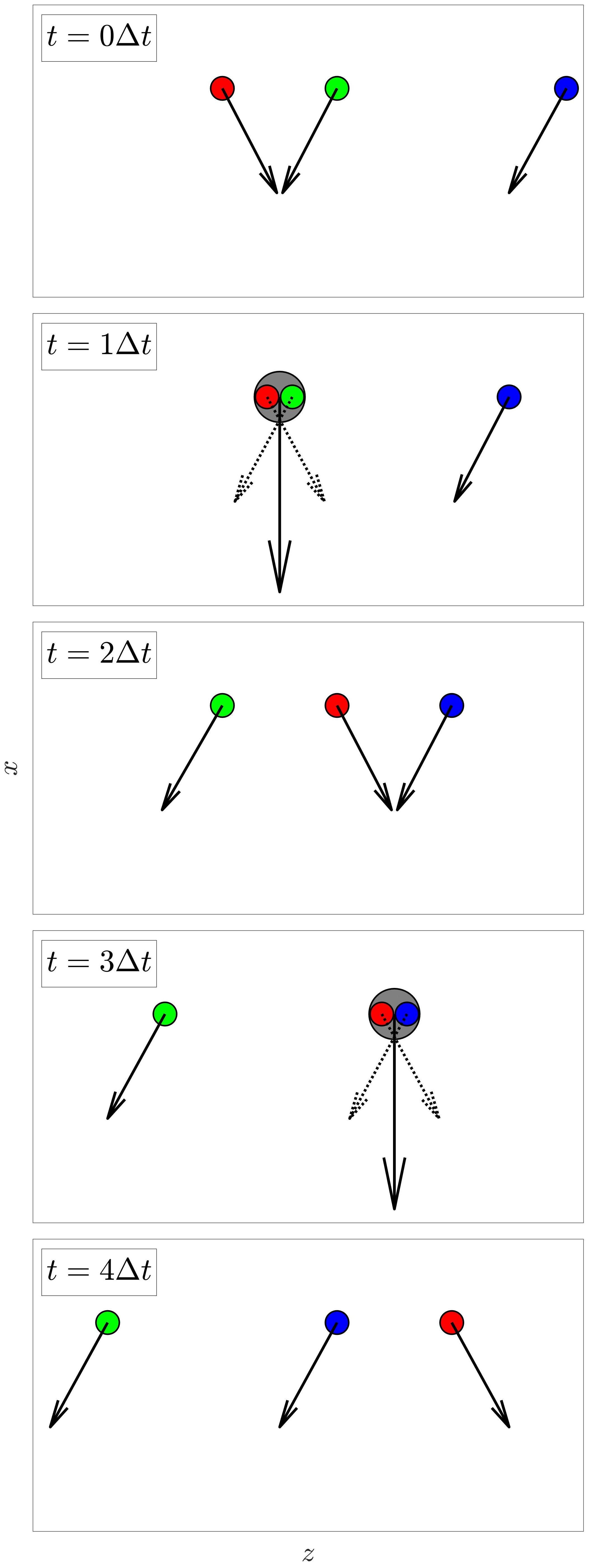}
    \caption{Schematic illustration of two SN contributions (initially at $\theta=120^\circ$ and $\theta=60^\circ$) interfering before complete decoupling, creating an apparent transient SN flux at angle $\theta=90^\circ$ and at two different spatial points.
    After full decoupling ($t=4\Delta t$), no particle is emitted at $\theta=90^\circ$.}
    \label{Fig:SM_illustration}
\end{figure}

The non-saturation at other angles arises from a transient interference effect illustrated schematically in Fig.~\ref{Fig:SM_illustration}: contributions to the SN density moving at different angles overlap as they decouple, producing an apparent (and spurious) SN flux at intermediate angles such as at $\theta = 90^\circ$.
This effect diminishes once the SNs have had sufficient time to separate spatially, but the present basis domain is too small for this to occur at all angles before $t_\mathrm{ref}$.
This illustration explains why the $\theta(\textbf{r},t)=90^\circ\pm 1^\circ$ region in Fig.~\ref{Fig:SM_dens_snapshots} is curved and why the yield $N_\theta^{(\mathrm{SN})}(t)$ for $\theta=90^\circ\pm 1^\circ$ in Fig.~\ref{Fig:SM_neut_time} has a non-monotone behaviour.

\section{Supplementary results: $^{239}\mathrm{Pu}(\mathrm{n}_{\mathrm{th}},\mathrm{f})$\label{Sec:SM_Pu240}}

\begin{figure}
    \centering
    \includegraphics[width=1.0\columnwidth]{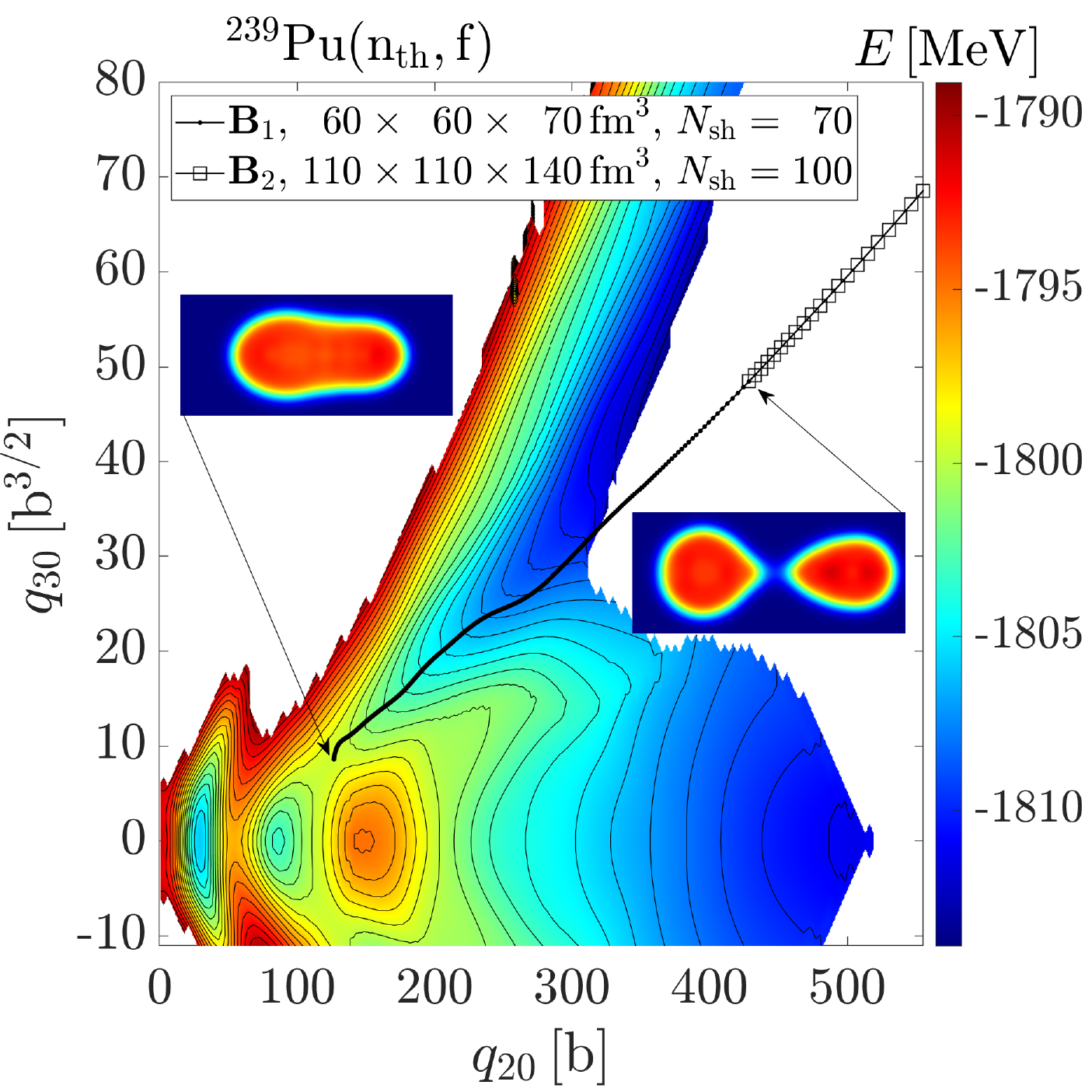}
    \caption{Potential energy surface (PES) of $^{240}\mathrm{Pu}$ with TDDFT trajectory initialized at the saddle point. Isoenergy contours spaced by 1 MeV; consecutive trajectory points separated by $5\,\mathrm{fm}/c$.
    Scission occurs at $t_\mathrm{sci}=2000\,\mathrm{fm}/c$ when the basis change from $\textbf{B}_1$ to $\textbf{B}_2$ is performed.}
    \label{Fig:SM_PES_Pu240}
\end{figure}

Fig.~\ref{Fig:SM_PES_Pu240} shows the $^{240}\mathrm{Pu}$ potential energy surface (PES) overlaid with the TDDFT trajectory starting at $q_{20}=126.8\,\mathrm{b}$ and $q_{30}=8.6\,\mathrm{b}^{3/2}$.

\begin{figure}
    \centering
    \includegraphics[width=1.0\columnwidth]{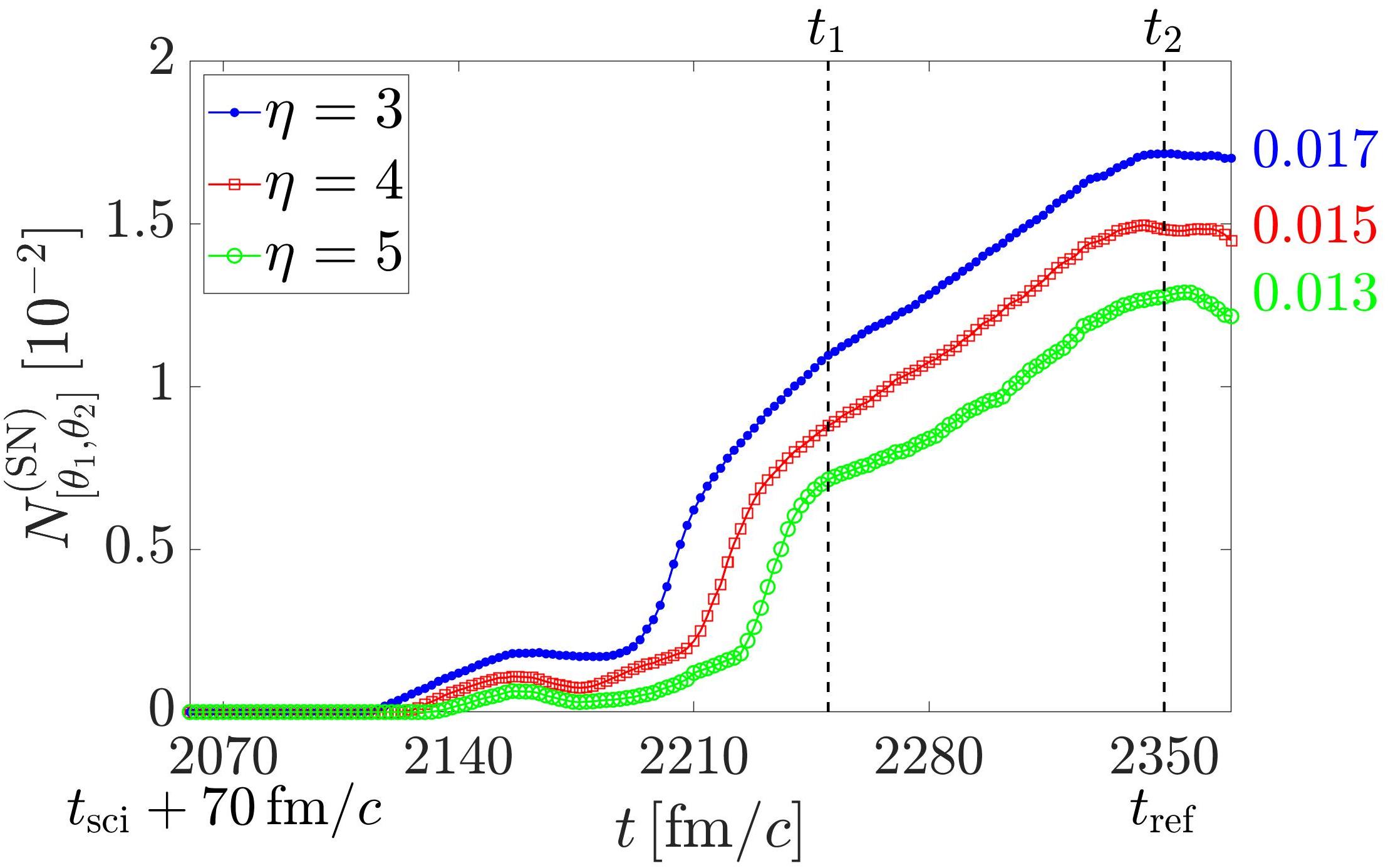}
    \caption{Time dependence of the number of SNs emitted into $[\theta_1,\theta_2]=[107.8^\circ,143.2^\circ]$ for $^{239}\mathrm{Pu}(\mathrm{n}_{\mathrm{th}},\mathrm{f})$, shown for $\eta=3$, 4, 5.
    Saturation values and the averaging interval $[t_1,t_2]$ used to compute the probability distribution $P^{(\mathrm{SN})}_{[\theta_1,\theta_2]}(E)$ are indicated.}
    \label{Fig:SM_Nsn_Pu240}
\end{figure}

Fig.~\ref{Fig:SM_Nsn_Pu240} shows the time dependence of the SN yield $N^{(\mathrm{SN})}_{[\theta_1,\theta_2]}$ in the angular interval $[\theta_1,\theta_2]=[107.8^\circ,143.2^\circ]$, demonstrating saturation for all three threshold values $\eta=3,4,5$.
Fig.~\ref{Fig:SM_Nsn_Pu240} also indicates the time interval $[t_1,t_2]$ over which the probability distribution $P^{(\mathrm{SN})}_{[\theta_1,\theta_2]}(E)$ is averaged.

\begin{figure}
    \centering
    \includegraphics[width=0.95\columnwidth]{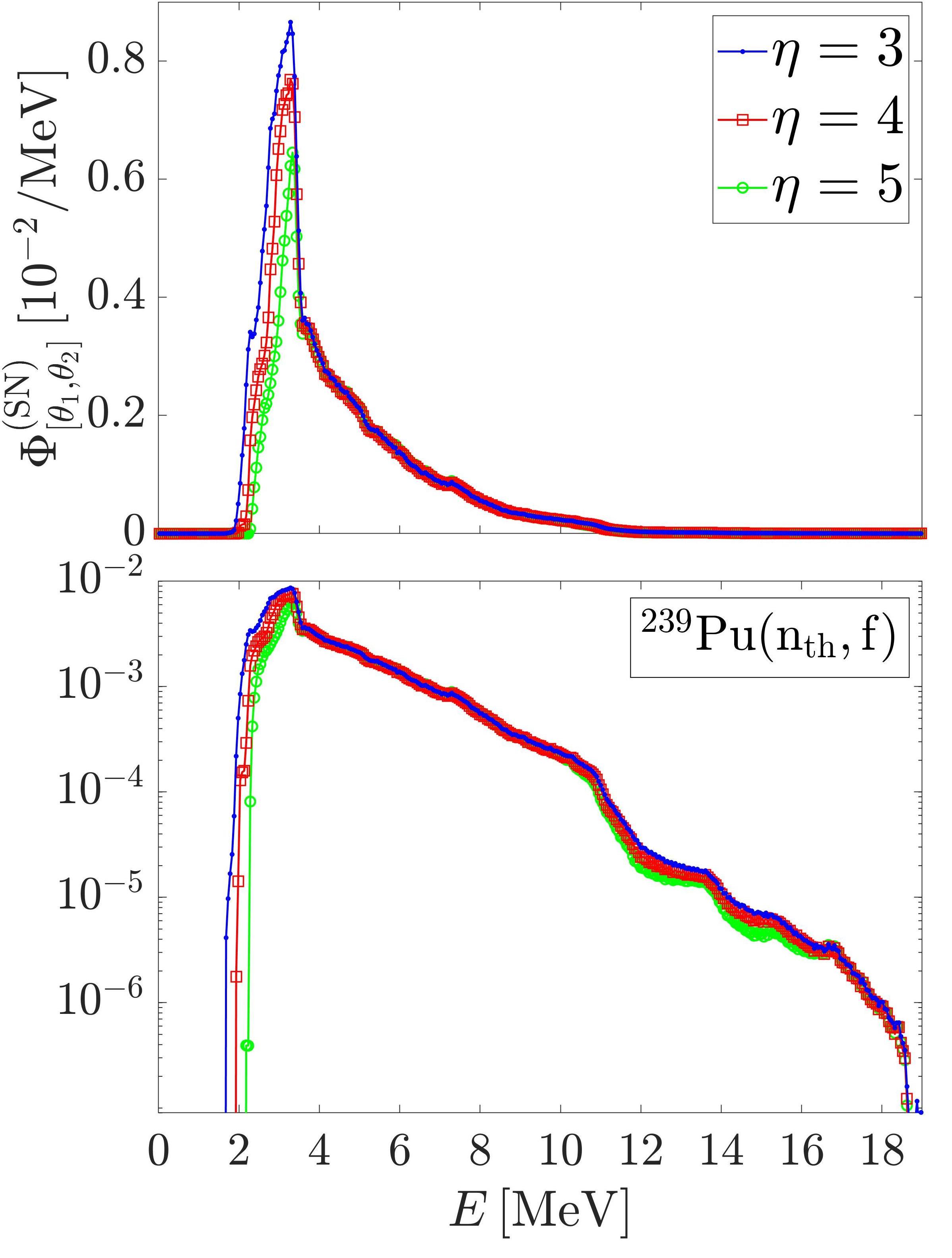}
    \caption{The SN energy distribution for $^{239}\mathrm{Pu}(\mathrm{n}_{\mathrm{th}},\mathrm{f})$ for $\eta=3$, 4, 5, is shown on linear (upper) and logarithmic (lower) scales.
    The distribution is insensitive to the value of $\eta$ above $3\,\mathrm{MeV}$; below $2\,\mathrm{MeV}$ it vanishes for all values of $\eta$.}
    \label{Fig:SM_Phisn_Pu240}
\end{figure}

Fig.~\ref{Fig:SM_Phisn_Pu240} shows the dependence of the SN energy distribution on the threshold parameter $\eta$.
Varying $\eta$ from 3 to 5 primarily affects the low-energy portion of the distribution (between 2 and 3 MeV), while leaving the peak position and high-energy tail essentially unchanged.
The distribution vanishes below $2\,\mathrm{MeV}$ for all values of $\eta$.
This confirms that the primary cutoff and exponential-tail features of the SN spectrum $\Phi^{(\mathrm{SN})}_{[\theta_1,\theta_2]}(E)$ are robust with respect to variations in the parameter $\eta$.

\section{Supplementary results: $^{252}\mathrm{Cf}(\mathrm{sf})$\label{Sec:SM_Cf252}}

\begin{figure}
    \centering
    \includegraphics[width=1.00\columnwidth]{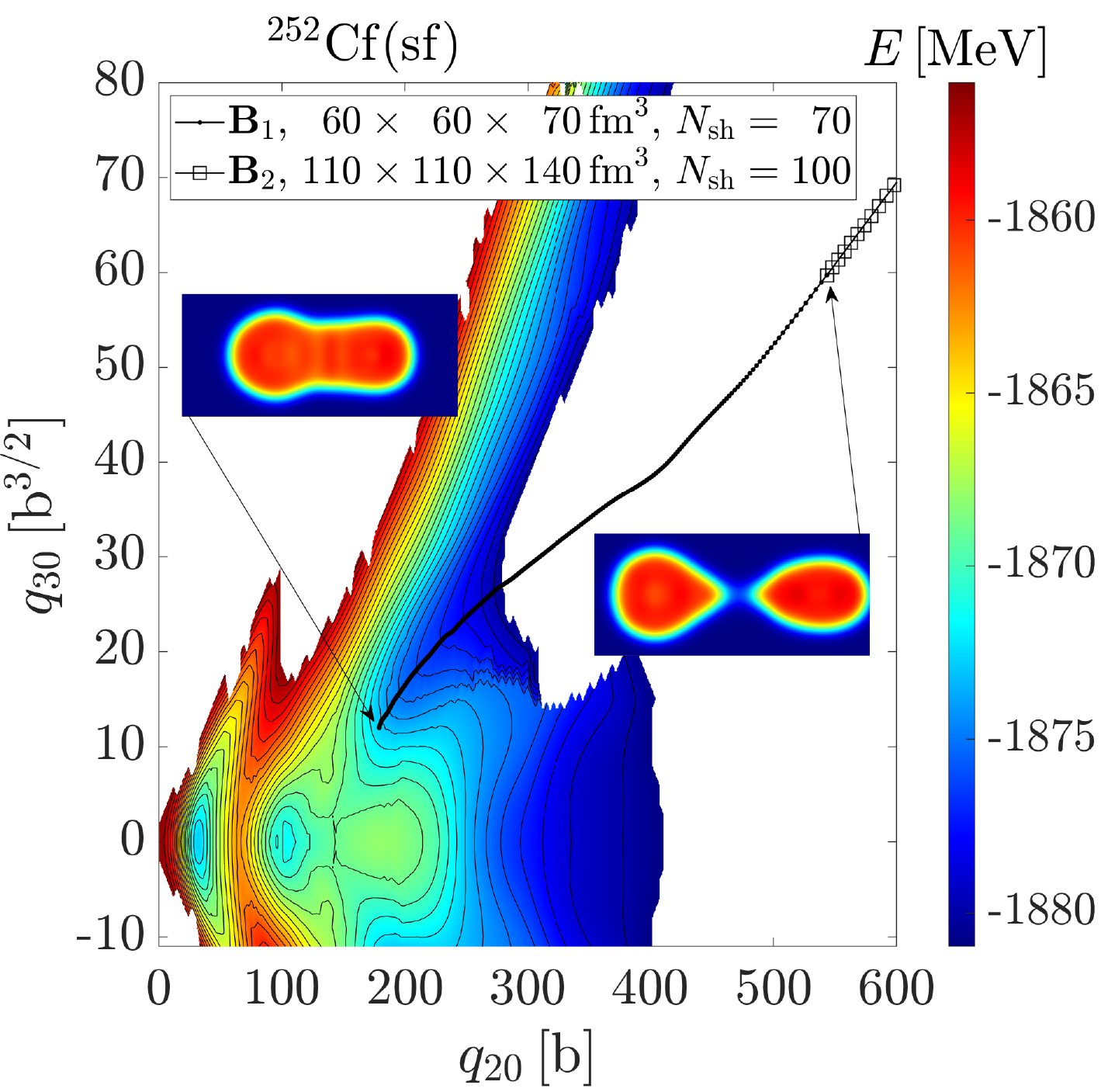}
    \caption{Same as Fig.~\ref{Fig:SM_PES_Pu240} but for $^{252}\mathrm{Cf}(\mathrm{sf})$, initialized at $q_{20}^0=179\,\mathrm{b}$, $q_{30}^0=12\,\mathrm{b}^{3/2}$. The energy of the configuration is the same as the ground state. Scission occurs at $t_\mathrm{sci}=1750\,\mathrm{fm}/c$.}
    \label{Fig:SM_PES_Cf252}
\end{figure}

Fig.~\ref{Fig:SM_PES_Cf252}
shows the $^{252}\mathrm{Cf}$ PES overlaid with the TDDFT trajectory.
The $^{252}\mathrm{Cf}(\mathrm{sf})$ spontaneous fission trajectory starts from a deformation beyond the outer saddle with the same HFB energy as the ground state.

\begin{figure}
    \centering
    \includegraphics[width=1.00\columnwidth]{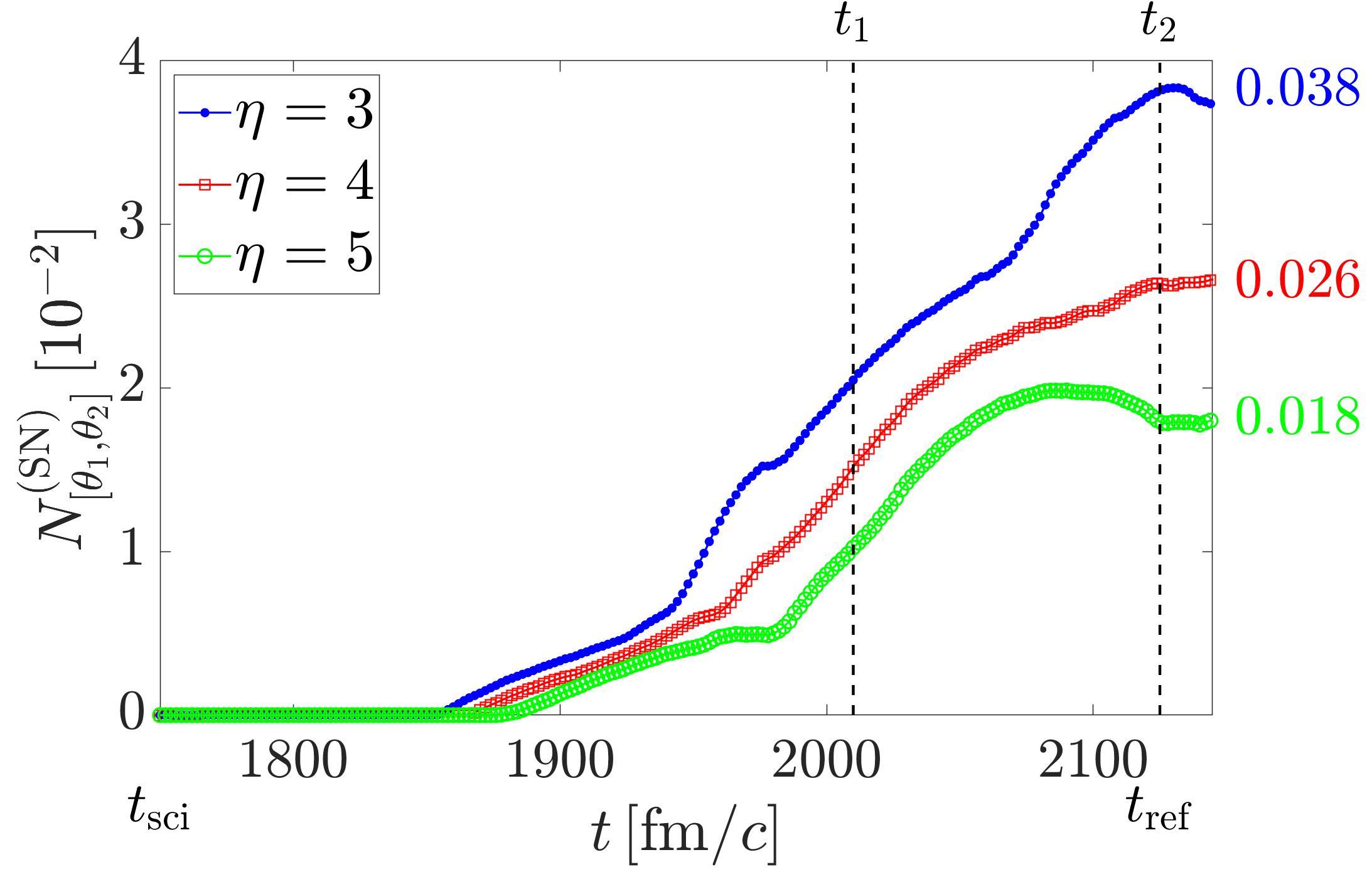}
    \caption{Same as Fig.~\ref{Fig:SM_Nsn_Pu240} but for $^{252}\mathrm{Cf}(\mathrm{sf})$.}
    \label{Fig:SM_Nsn_Cf252}
\end{figure}

The angular stability analysis for $^{252}\mathrm{Cf}(\mathrm{sf})$ yields the same reliable extraction range ($110^\circ \lesssim \theta \lesssim 150^\circ$) as for $^{239}\mathrm{Pu}(\mathrm{n}_{\mathrm{th}},\mathrm{f})$.
Fig.~\ref{Fig:SM_Nsn_Cf252} shows the same saturation property of $N^{(\mathrm{SN})}_{[\theta_1,\theta_2]}$ as in Fig.~\ref{Fig:SM_Nsn_Pu240}.

\section{Experimental data processing\label{Sec:SM_expdata}}

\begin{figure}
    \centering
    \includegraphics[width=\columnwidth]{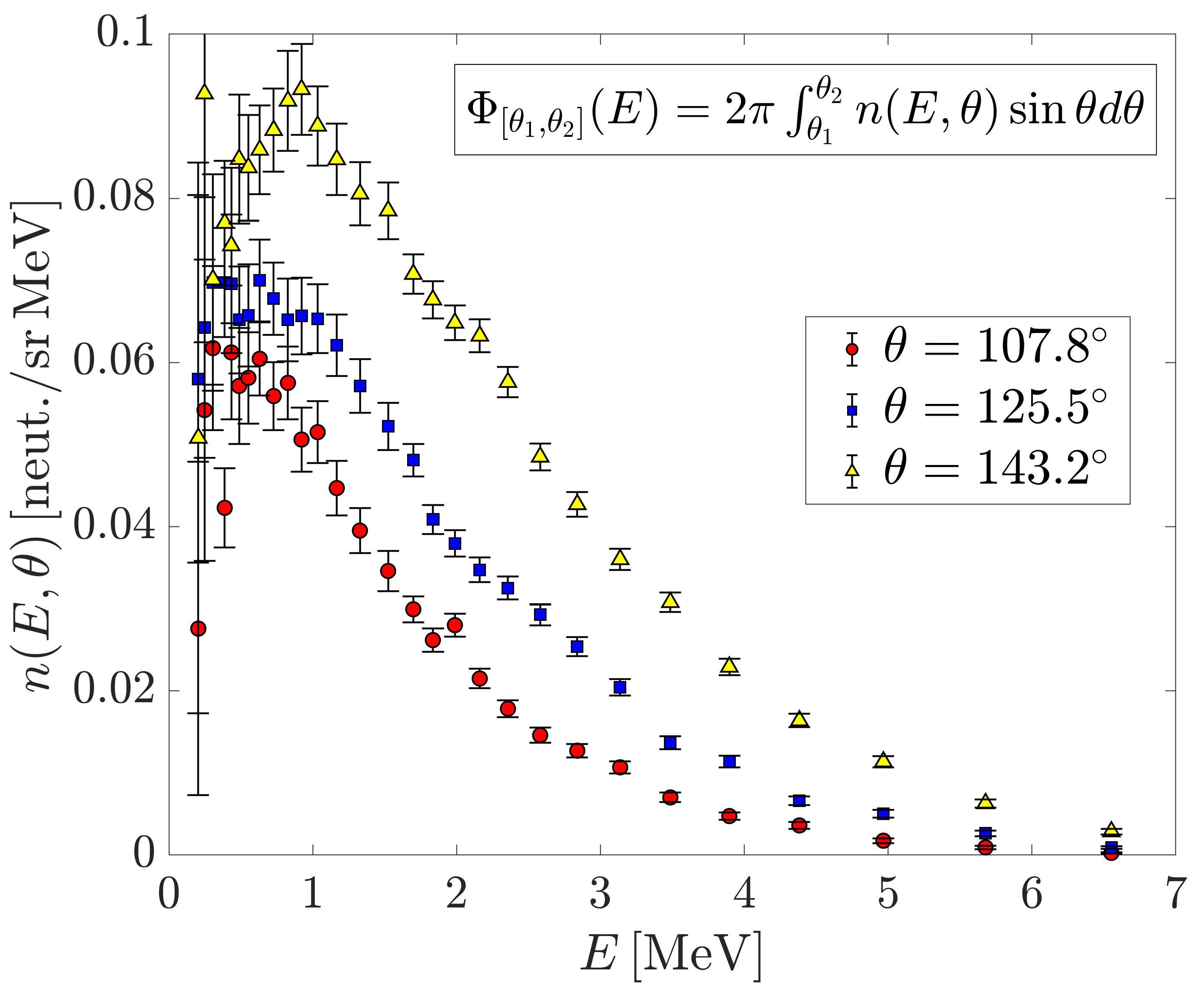}
    \caption{Experimental data from Ref.~\cite{Vorobyev2018ScissionPu239} for the three angles used in the present analysis.
    }
    \label{Fig:SM_expdata}
\end{figure}

\textcite{Vorobyev2018ScissionPu239,Vorobyev2017ScissionCf252} tabulate measurements of $n(E,\theta)$, the number of prompt fission neutrons emitted per unit energy and solid angle, for 11 discrete angles $\theta_j$ and 33 discrete energies $E_i$.
Because DFT calculations are currently only reliable for $110^\circ \lesssim \theta \lesssim 150^\circ$, results in this study are compared to the three measured angles closest to the aforementioned interval,
$\theta^{(j)} = 107.8^\circ + (j-1)\times 17.7^\circ$, $j=1,2,3$.

To improve the signal-to-noise ratio in the $0$--$2\,\mathrm{MeV}$ fit region, these three angles are integrated over using the trapezoidal rule:
\begin{equation}
    \Phi_{[\theta_1,\theta_2]}(E)
    =
    2\pi \Delta\theta
    \sum_{j=1}^{3}
    c_j\, \sin\theta^{(j)}\, n(E,\theta^{(j)}),
\end{equation}
with $c_1=c_3=\frac{1}{2}$, $c_2=1$, and $\Delta\theta = 17.7^\circ$.
The propagated uncertainty is
\begin{equation}
    \sigma_{[\theta_1,\theta_2]}(E)
    =
    2\pi \Delta\theta
    \sqrt{
    \sum_{j=1}^{3}
    \left[
        c_j\,\sin\theta^{(j)}\,
        \sigma_{n(E,\theta^{(j)})}
    \right]^2 }.
\end{equation}
Three individual datasets from Ref.~\cite{Vorobyev2018ScissionPu239} are shown in Fig.~\ref{Fig:SM_expdata}.
The final experimental data used in Eq.~\eqref{Eq:chi2}
is obtained as $\Phi_{[\theta_1,\theta_2]}^{(i)}\pm \sigma^{(i)}_{[\theta_1,\theta_2]}  = \Phi_{[\theta_1,\theta_2]}(E_i) \pm \sigma_{[\theta_1,\theta_2]}(E_i)$.

Analogous processing is performed for $^{252}\mathrm{Cf}(\mathrm{sf})$ using the data of Ref.~\cite{Vorobyev2017ScissionCf252}.

\bibliography{zotero_output,books,others}